# AGNOSTIC SEQUENTIAL RATIONALITY

Igal Milchtaich, Department of Economics, Bar-Ilan University, Israel
igal.milchtaich@biu.ac.il https://sites.google.com/view/milchtaich

*Agnostic sequential equilibrium* (ASE) is a refinement of sequential equilibrium that does not force on the players a single, arbitrary belief system. In addition, whereas sequential equilibrium assumes the players' beliefs to be fully consistent (a notion that is based on perturbations of strategies), ASE employs a novel, simpler and local concept of *strong consistency* between strategy profiles and off-equilibrium beliefs, which is applicable to a large class of dynamic games, including games with a continuum of actions. In the last respect, the new solution concept is similar to perfect Bayesian equilibrium. It is shown that a strategy profile in an imperfect-information extensive-form game with perfect recall is an ASE precisely when it is a sequential equilibrium with *every* fully consistent belief system. ASE is generalized by the set-valued solution concept of agnostic sequential *poly*equilibrium, which allows leaving the players' actions in some information sets partially or completely unspecified.

## 1 Introduction

In dynamic games with perfect information, the idea that the players' choices of strategies should be rational also off-equilibrium, that is, at decision nodes that are not actually reached, is captured by the notion of subgame perfect equilibrium. In particular, this refinement of Nash equilibrium excludes non-credible threats: actions that are detrimental to the actor and a rational player would therefore not carry out. In games with imperfect information, an action taken at a particular information set $U$ may be beneficial or detrimental depending on the specific node at which it is carried out. However, by definition of information set, the player does not know which of the nodes in $U$ was reached. Therefore, the merit of choosing a particular action can be assessed only with respect to particular beliefs about the history of play, which are expressed by a probability distribution over the nodes in $U$. An obvious problem is that if $U$ is an off-equilibrium information set, then the prior probability of every sequence of actions leading to it is zero, and so the beliefs in question cannot be derived from Bayes' rule as they refer to an event (namely, reaching $U$) that should not have happened if the players adhered to their equilibrium strategies.

A standard solution to this indeterminacy problem is to provide the beliefs as part of the solution concept. Thus, a solution is a pair $(\mu, x)$, called an *assessment*, where $x$ is a strategy profile and $\mu$ is a *belief system* that specifies a probability distribution over the nodes in every information set $U$ of every player. The probability assigned to a set of nodes $V \subseteq U$ is denoted $\mu(V)$ (and so $\mu(U) = 1$). This approach is employed by the sequential equilibrium solution concept (Kreps and Wilson 1982) as well as by the more general, but somewhat nebulous, notion of perfect Bayesian equilibrium (a precise version of which — albeit in a special setting — is given by Fudenberg and Tirole 1991).[1]

Sequential equilibrium, perfect Bayesian equilibrium and similar solution concepts require the players' strategies in $x$ to be *sequentially rational* under $\mu$, meaning that the strategy $x_i$ of each player $i$ is optimal in the continuation game starting at each of the player's information sets $U$, given the other players' strategies and $i$'s beliefs at $U$ about the events that preceded the arrival there, which are specified by $\mu$. They also require these hypotheses, or conjectures, about the history of play to be reasonable. Different solution concepts have different definitions of "reasonable" hypotheses and different requirements as to how they should reflect the players' equilibrium strategies. However, they all leave at least some leeway when off-equilibrium information sets are concerned, which means that beliefs are often effectively *chosen* much like strategies are chosen. However, unlike the choice of strategies, which needs to be justified in the sense of satisfying sequential rationality, off-equilibrium beliefs may be largely arbitrary. This is a divergence from the common view that equilibrium represents a self-enforcing convention, as the beliefs at off-equilibrium information sets are neither self-enforcing nor can they be described as a convention, because these sets are actually never reached in equilibrium. The specification of actions that are justifiable only by the beliefs they are bundled with may also be difficult to reconcile with the prescriptive view of equilibrium.

As an example, consider the equilibrium shown in Figure 1a. It is a sequential (and perfect Bayesian) equilibrium, because player 2's choice of $r$ rather than $l$ is justified by *some* belief, namely, that if player 1 deviated from his equilibrium strategy of $Out$, it was by playing $R$ rather than $L$. However, all alternative beliefs of player 2 at the (off-equilibrium) information set $U$ are arguably as reasonable as this one. But if these beliefs, in particular, a belief that 1 played $L$, are not excluded, then a choice of $l$ by player 2 also cannot be excluded. Depending on the value of the parameter $\alpha$, this in turn may call into question player 1's choice of $Out$.

The concept of agnostic sequential equilibrium proposed in this paper differs from sequential and perfect Bayesian equilibrium in that it does not allow specifying actions in off-equilibrium information sets (for example, $r$ in Figure 1a) that can be justified only under particular, arbitrary beliefs about the history of play. The choice of such actions entails the exclusion of one or more alternative actions that under different but equally reasonable beliefs are actually better. The concept is based on the idea that an action may be excluded only if some other action at the same informtion set $U$ gives the acting player the same or higher payoff under all beliefs in $U$ that are consistent with the strategy profile. Key for implementing this idea is pinning down the appropriate meaning of consistency in this context.

---

[1] A natural extension of the terminology is to refer to a *strategy profile* $x$ as a sequential or perfect Bayesian equilibrium when there is *some* belief system $\mu$ such that the assessment $(\mu, x)$ is so.

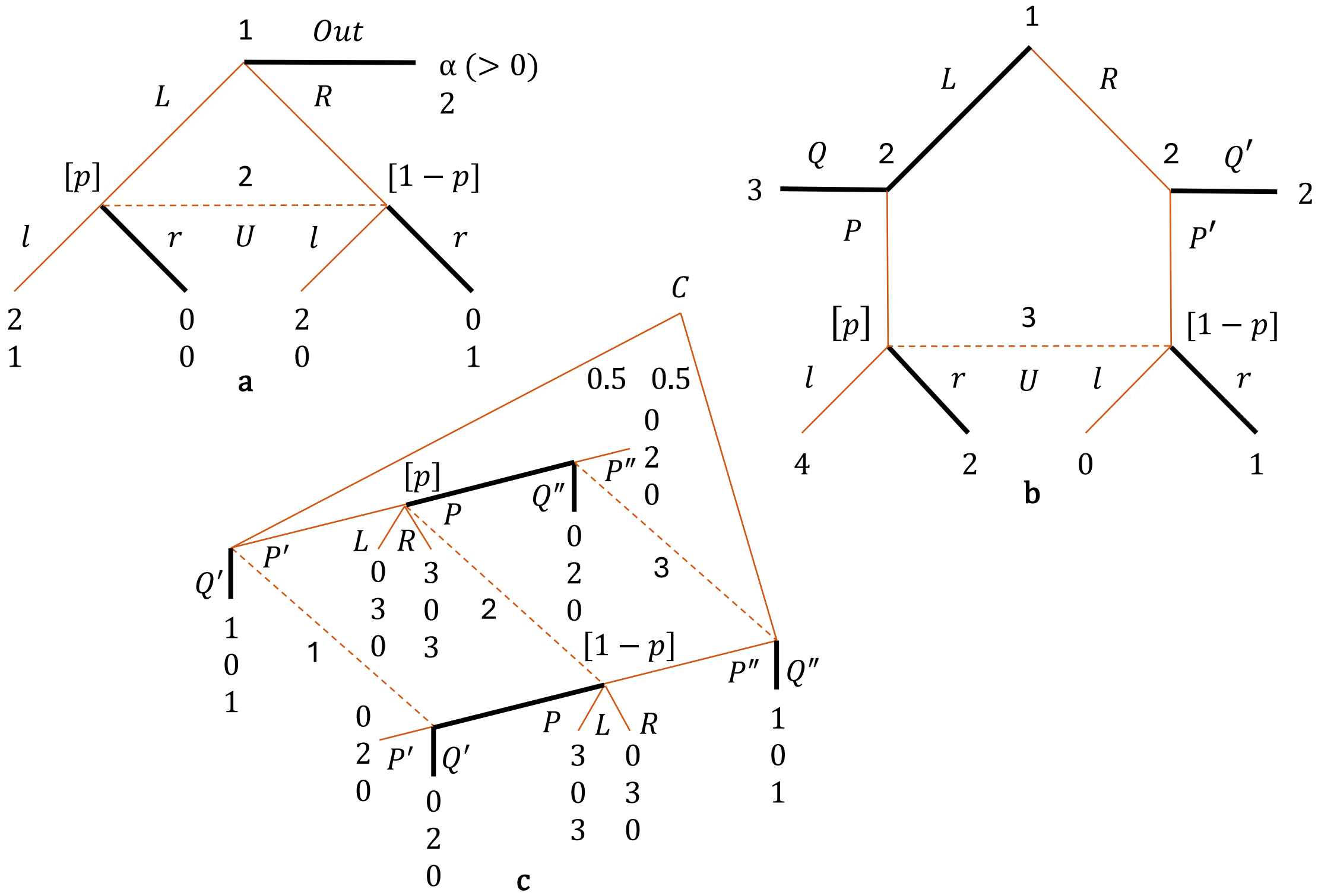

**Figure 1 Sequential equilibria that are not agnostic (black lines). a Player 2's choice of $r$ is justified by the belief $p = 0$ but not by $p = 1$. b Player 3's choice of $r$ is justified by $p = 0$ but not by $p = 1$. In this game, the players have identical payoffs. c Player 2's choice of $P$ is justified only if $1/3 \leq p \leq 2/3$. In this game, $C$ is a chance move that determines whether the players play in the order 1-2-3 or 3-2-1.**

Section 2 introduces the notion of strongly consistent beliefs. This concept reflects the idea that an explanation for the arrival at an off-path information set $U$ should invoke a minimal (with respect to inclusion) set of deviations from the players' strategy profile — a parsimonious explanation. The collection of all strongly consistent beliefs is shown to be a subset of the fully consistent beliefs, which are those used in sequential equilibrium and arise from the very different idea of completely mixed trembles. At the same time, all fully consistent beliefs are convex combinations of strongly consistent ones.

A conceptual advantage of strongly consistent beliefs over those used in sequential and perfect Bayesian equilibrium is that only the former beliefs are guaranteed to be structurally consistent (Kreps and Ramey 1987). That is, they are always consistent with the assumption that, after $U$, there will be no more deviations from the original strategy profile. This assumption is implicit in the definition of sequential rationally but cannot always be upheld in a sequential equilibrium.

The formal definition of agnostic sequential equilibrium is presented in Section 3. This definition is somewhat unusual in that it does not explicitly mentions beliefs. Instead, strongly consistent beliefs are implicit in the definition's reference to all strategy profiles that arise from modifying the original profile only at a set of actions that is minimally sufficient for reaching the information set under consideration.

The relations pointed to above between strongly and fully consistent beliefs imply a very sharp relation between the corresponding solution concepts. Namely, a strategy profile is an agnostic sequential equilibrium if and only if it is a sequential equilibrium with respect to *all* fully consistent beliefs. This strong requirement entails that such an equilibrium does not always exist. (More on this below.)

Because agnostic sequential equilibrium is not based on trembles, it is applicable without modification to games with a continuum of actions, which makes it comparable to perfect Bayesian equilibrium. Agnostic sequential rationality does not however overcome certain known conceptual difficulties that may arise when dealing with this kind of games (Myerson and Reny 2015, 2020). As explained in Section 4, the difficulties stem from the fact that, with a continuum of available moves, arrival at a zero-probability information set may be possible, or even guaranteed, without anyone deviating.

The principle of parsimony underlying strong consistency of beliefs is simple but crude. It may — and should — be refined to better reflect how players can be expected to reason under particular circumstances. As is the case for sequential and perfect Bayesian equilibria, it is difficult to point to a refinement that would make sense in all settings. However, it is possible to identify certain broad desiderata. An obvious one is the requirement that beliefs about other players' past deviations from their strategies should be congruent with the strategic interests of those players. (Like sequential equilibrium, the beliefs underlying agnostic sequential equilibrium only reflect the structure of the game tree, not the payoffs.) As shown in Section 5, adopting this requirement may have two different outcomes. One possible outcome is invalidation. Strongly consistent beliefs may have to be discarded because, for example, they imply that a strictly dominated strategy was used while other beliefs do not imply this. The other, more constructive, outcome is pruning: the set of strongly consistent beliefs shrinks. Such shrinkage *enlarges* the set of agnostic sequential equilibria, which is the opposite of the effect that a similar refinement has on sequential and perfect Bayesian equilibria. Consequently, the gap between these solution concepts and agnostic sequential equilibrium may narrow. It is shown that, in the Beer-Quiche game for example, the gap may disappear altogether.

The problem of the possible nonexistence of agnostic sequential equilibrium can be formally solved by extending this solution concept to a set-valued one. Section 6 introduces agnostic sequential polyequilibrium, a set-valued extension that is a special case of the general polyequilibrium concept (Milchtaich 2019). As a set-valued solution specifies a set of strategy profiles, the extension may seem to only shift the problem elsewhere: the problem of nonexistence is traded for that of assigning a predictive meaning to the set. However, a set of strategy profiles does have a clear predictive content, which is the collection of properties common to all its elements. Depending on the particular polyequilibrium examined, this content, called polyequilibrium results, may be rich or poor. An example of the former is provided by the game in Figure 1a, with $\alpha > 2$. This game has no agnostic sequential equilibrium but has a nontrivial agnostic sequential polyequilibrium that consists of all strategy profiles where player 1's strategy is (the strictly dominant one) $Out$. Although player 2's strategy is not specified, the polyequilibrium has real content. Indeed, it completely specifies the game's outcome.

Unless where otherwise stated, the discussion below assumes that the dynamic game under consideration is an *extensive form game*, that is, one that can be described by a finite game tree, possibly with chance nodes (where, without loss of generality, all outcomes are assumed to have positive probability). It also assumes perfect recall, which in particular means that each player's collection of information sets is partially ordered by precedence. Throughout, 'strategy' may refer to either *pure* strategy, which prescribes a single action at each of the player's information sets, or *behavior* strategy, which prescribes a probability distribution over actions in each information set. The former is a special case of the latter, with all probabilities 0 or 1. Whether all behavior strategies or only pure ones can be used is viewed as part of the game's specification. All the results in the paper hold for both cases, and any reference to (an unqualified) 'strategy' may be understood in either way. An additional notational convention (which involves no loss of generality) is that actions at different information sets are always labeled differently, so that an action's label uniquely identifies the information set at which it is playable and the identity of the acting player.

# 2 Consistent Beliefs

A minimal consistency requirement for an assessment $(\mu, x)$ is *weak consistency* of the assessment (or of the belief system $\mu$ *with* the strategy profile $x$), which means that the probabilities specified by $\mu$ coincide with those derived from $x$ using Bayes' rule (in other words, with the conditional probabilities) whenever possible. Specifically, for an information set $U$ and a subset $V \subseteq U$, denote by $\mathbb{P}_x(V)$ the (prior, unconditional) probability that, under $x$, one of the nodes in $V$ is reached. If the information set is reached with positive probability (that is, $\mathbb{P}_x(U) > 0$), then the conditional probability that $V$ is reached, given that $U$ is reached, is $\mathbb{P}_x(V \mid U) = \mathbb{P}_x(V) \,/\, \mathbb{P}_x(U)$. The weak consistency requirement is that, for every information set $U$,

$$\text{if } \mathbb{P}_x(U) > 0, \text{ then } \mu(V) = \mathbb{P}_x(V \mid U), \qquad V \subseteq U. \tag{1}$$

A standard strengthening of (1), which may be called *augmented weak consistency* at the information set $U$, is the requirement that, for every strategy $x_i'$ of the player $i$ acting at $U$, a condition similar to (1) holds with $x$ replaced with the strategy profile $x \mid x_i'$ obtained by replacing $i$'s strategy $x_i$ with $x_i'$. The additional requirement has a bite only if the information set $U$ is *reachable* for player $i$ under $x$, in the sense that $\mathbb{P}_{x|x_i'}(U) > 0$ for some strategy $x_i'$. The reason the requirement is sensible is that, since player $i$ has a perfect recall of his actions, all nodes in $U$ are preceded by the same sequence of $i$'s actions and therefore the relative probabilities that the nodes are reached only depend on the probabilities of the other players' actions. Thus, for all strategies $x_i'$ and $x_i''$ of player $i$,

$$\text{if } \mathbb{P}_{x|x_i'}(U), \mathbb{P}_{x|x_i''}(U) > 0, \text{ then } \mathbb{P}_{x|x_i'}(V \mid U) = \mathbb{P}_{x|x_i''}(V \mid U), \qquad V \subseteq U. \tag{2}$$

Solution concepts that involve a single, specific belief system $\mu$ usually impose on it also certain *internal consistency* requirements, which express the idea that beliefs at different off-equilibrium information sets should not only reflect the players' strategy profile but also represent a coherent hypothesis about the players' deviations from it. In particular, at an information set that follows another information set of the same player, beliefs should be derived from the beliefs at the earlier set whenever possible. This requirement is formally expressed by the *preconsistency* condition (Hendon et al. 1996; Perea 2002), which is based on Fudenberg and Tirole's (1991) notion of reasonable assessment. Internal consistency between beliefs at information sets belonging to different players is guaranteed by the stronger condition of *full consistency* of the assessment $(\mu, x)$ (or of the belief system $\mu$ with the strategy profile $x$). The condition requires the assessment to be the (pointwise) limit of some sequence of weakly consistent assessments $(\mu^k, x^k)_{k=1}^{\infty}$ where each $x^k$ is a *completely mixed* strategy profile, in the sense that it assigns positive probability to every action at every information set (which entails that $\mu^k$ is uniquely determined by the weak consistency requirement). A *sequential equilibrium* (Kreps and Wilson 1982) is a fully consistent assessment $(\mu, x)$ such that $x$ is sequentially rational under $\mu$.

Agnostic sequential equilibrium does not specify a single belief system, which renders the whole internal consistency requirement moot. This brings about a considerable simplification, since consistency is narrowed down to the "local" condition that beliefs are reconcilable with the strategy profile $x$ at each individual information set $U$. If, under $x$, the probability that $U$ is reached is positive, this local consistency requirement is simply the weak consistency condition expressed by (1). However, if the probability is zero, then weak consistency does not specify any beliefs at $U$. Nevertheless, the player $i$ acting at $U$ may actually *know* something about the history of play there. In particular, if all nodes in $U$ are preceded by a particular action $a$ of some player $j$, then player $i$ knows that player $j$ took that action. (By the perfect-recall assumption, this knowledge applies in particular whenever $j = i$ and $a$ is one of $i$'s own actions leading to $U$.) This is so even if the probability specified to action $a$ by the strategy profile $x$ is 0. If $U$ was reached, then the specification was obviously not followed.

**Notation** For a strategy profile $x$ and a set of actions $\mathcal{A}$ (of one or more players), with each action playable at a different information set, the strategy profile obtained from $x$ by assigning probability 1 to every action in $\mathcal{A}$ (and probability 0 to all other actions in the respective information set) is denoted $x^{\mathcal{A}}$.

For a given information set $U$, let $\mathcal{A}$ be the set of all actions $a$ that precede all the nodes in $U$ and are specified probability zero by $x$. If $U$ is reached, then these actions were necessarily taken. For the player acting in $U$, it only remains to speculate about the other players' behavior at information sets that do not involve actions in $\mathcal{A}$. The simplest hypothesis is that they adhere to $x$ there. In other words, the hypothesis effectively replaces $x$ with the strategy profile $x^{\mathcal{A}}$ specifying that every action in $\mathcal{A}$ is taken with probability 1 rather than 0.[2]

[2] Probability 1 here means that the action was in fact taken, not that the acting player $j$ necessarily *meant* to play it for sure. However, 1 could be replaced by any other positive probability, as doing so for one or more actions in $\mathcal{A}$ has no effect on the beliefs at $U$.

If, under $x^{\mathcal{A}}$, the information set $U$ is reached with positive probability, then this hypothesis yields a unique probability distribution on $U$, which is the conditional distribution $\mathbb{P}_{x^{\mathcal{A}}}(\cdot| U)$. This distribution, which arguably represents the only beliefs at $U$ that are consistent with $x$, will be said to be strongly consistent with the strategy profile $x$. However, the probability that $U$ is reached may be zero even under $x^{\mathcal{A}}$, and in this case, reaching $U$ indicates that there is at least one action outside $\mathcal{A}$ that is precluded (i.e., assigned probability zero) by $x$ but was nevertheless taken. A natural approach in this case is to enlarge the set $\mathcal{A}$ by adding to it one or more precluded actions, each playable at a different information set, in such a way that $\mathcal{A}$ becomes minimally sufficient for reaching $U$ under $x$, in the sense that $\mathbb{P}_{x^{\mathcal{A}}}(U) > 0$ but $\mathbb{P}_{x^{\mathcal{B}}}(U) = 0$ for all $\mathcal{B} \subsetneq \mathcal{A}$. Such an enlargement is not unique. There may be several sets of actions $\mathcal{A}$ that are minimally sufficient for reaching $U$, yielding different (indeed, mutually singular; see (ii) in Lemma 1 below) conditional distributions $\mathbb{P}_{x^{\mathcal{A}}}(\cdot| U)$. These distributions, which represent alternative hypotheses about the actions that preceded the arrival at $U$, will *all* be referred to as strongly consistent with the strategy profile $x$.

**Definition 1** A set of actions $\mathcal{A}$, each playable at a different information set, is *minimally sufficient* for reaching an information set $U$ under a strategy profile $x$ if (i) $U$ is reached with positive probability under $x^{\mathcal{A}}$, and (ii) no proper subset of $\mathcal{A}$ has a similar property. A probability distribution over the nodes in $U$ is *strongly consistent* with $x$ if it coincides with the conditional distribution $\mathbb{P}_{x^{\mathcal{A}}}(\cdot| U)$ for some minimally sufficient set $\mathcal{A}$.

Note that if $x$ in Definition 1 is a pure-strategy profile and there are no chance moves in the game, then all strongly consistent beliefs are degenerate; each of them assigns probability 1 to some node in $U$. For example, in the game in Figure 1a, the two strongly consistent beliefs for player 2 are $p = 1$ and $p = 0$, corresponding to $\mathcal{A} = \{L\}$ and $= \{R\}$ respectively. The same is true for player 3 in the game in Figure 1b, where both $\{P\}$ and $\{R, P'\}$ are minimally sufficient for reaching the information set $U$.[3]

Strong consistency at an information set implies weak consistency there, because if an information set $U$ is reached with positive probability under a strategy profile $x$, then the unique minimally sufficient set of actions for reaching $U$ is $\mathcal{A} = \emptyset$, for which $x^{\mathcal{A}} = x$. It also implies augmented weak consistency, because if $U$ is reachable for the player $i$ acting there, then the unique minimally sufficient set $\mathcal{A}$ consists of all actions of $i$ that are assigned probability zero by $x_i$ and precede (all the nodes in) $U$, and so $x^{\mathcal{A}} = x \mid x_i'$ for some strategy $x_i'$. A less obvious fact is that strong consistency furthermore implies the local version of full consistency. This fact is established by the following lemma, which moreover shows that, at every information set $U$,

$$\mathfrak{B}^{\mathrm{S}} \subseteq \mathfrak{B}^{\mathrm{F}} \subseteq \operatorname{conv} \mathfrak{B}^{\mathrm{S}}, \tag{3}$$

where $\mathfrak{B}^{\mathrm{S}}$ is the set of all strongly consistent beliefs, $\operatorname{conv}\ \mathfrak{B}^{\mathrm{S}}$ is the convex hull of this set, and $\mathfrak{B}^{\mathrm{F}}$ is the set of all beliefs at $U$ arising from belief systems that are fully consistent with $x$.

[3] A hypothesis involving earlier deviations by two players may intuitively seem more strained than one involving a single deviation (or, perhaps, even multiple deviations) by a single player. However, the minimality condition presented above does not take a stand on this issue. It represents a conservative approach, and only precludes patently unnecessary assumptions.

In general, both inclusions in (3) may be proper.[4] It follows from these inclusions that a sufficient condition for $\mathfrak{B}^{\mathrm{S}}$ and $\mathfrak{B}^{\mathrm{F}}$ to coincide is that either set is a singleton, and a necessary and sufficient condition for $\mathfrak{B}^{\mathrm{F}}$ and $\operatorname{conv}\mathfrak{B}^{\mathrm{S}}$ to coincide is that the former is a convex set. Another corollary of (3) is that a degenerate belief is an element of $\mathfrak{B}^{\mathrm{S}}$ if and only if it is in $\mathfrak{B}^{\mathrm{F}}$.

**Lemma 1** For a strategy profile $x$ and an information set $U$, the collection $\mathfrak{A}$ of all sets of actions that are minimally sufficient for reaching $U$ under $x$ has the following properties:

(i) For each $\mathcal{A} \in \mathfrak{A}$ there is some node $u \in U$ such that $\mathcal{A}$ consists of all actions that precede $u$ and are assigned probability 0 by $x$.

(ii) The conditional distributions $\{\mathbb{P}_{x^{\mathcal{A}}}(\cdot \mid U)\}_{\mathcal{A}\in\mathfrak{A}}$ have pairwise disjoint supports.

(iii) For each $\mathcal{A} \in \mathfrak{A}$, the corresponding conditional distribution coincides with the distribution on $U$ specified by some belief system $\mu$ that is fully consistent with $x$:

$$\mathbb{P}_{x^{\mathcal{A}}}(V \mid U) = \mu(V), \qquad V \subseteq U. \tag{4}$$

(iv) For every belief system $\mu$ that is fully consistent with $x$, the distribution on $U$ specified by $\mu$ is a convex combination of the distributions $\{\mathbb{P}_{x^{\mathcal{A}}}(\cdot \mid U)\}_{\mathcal{A}\in\mathfrak{A}}$:

$$\mu(V) = \sum_{\mathcal{A}\in\mathfrak{A}} \lambda_{\mathcal{A}}\, \mathbb{P}_{x^{\mathcal{A}}}(V \mid U), \qquad V \subseteq U \tag{5}$$

for some (unique, in view of (ii)) nonnegative coefficients $\{\lambda_{\mathcal{A}}\}_{\mathcal{A}\in\mathfrak{A}}$ that sum up to 1.

*Proof.* For $\mathcal{A} \in \mathfrak{A}$, with cardinality $|\mathcal{A}|$ $(\geq 0)$, consider any node $u \in U$ with $\mathbb{P}_{x^{\mathcal{A}}}(\{u\}) > 0$.

Every action that precedes $u$ and is assigned probability 0 by $x$ must clearly be in the set $\mathcal{A}$. By the minimal sufficiency condition, $\mathcal{A}$ cannot include any other actions. This proves assertion (i). To establish (ii), it has to be shown that every $\mathcal{A}' \neq \mathcal{A}$ in $\mathfrak{A}$ satisfies $\mathbb{P}_{x^{\mathcal{A}'}}(\{u\}) = 0$. For this, it suffices to note that the actions in $\mathcal{A} \setminus \mathcal{A}'$ are assigned probability zero by $x^{\mathcal{A}'}$.

To prove (iii), let $z$ be some fixed completely mixed strategy profile and, for $0 < \epsilon < 1/2$, let $x^{\epsilon}$ be the strategy profile that, at each information set, assigns the following

[4] Consider, for example, a game where player 5 acts after players 1 through 4 choose $L$ or $R$, but he only knows how many of these players chose each action. Suppose that $x$ specifies that everyone chooses $R$, and consider the off-path information set $U$ where player 5 is informed that only two of his predecessors actually did so. The set $\mathfrak{B}^{\mathrm{S}}$ consists of the six degenerate distributions: those that assign the value 1 to the probability $p_{ij}$ that players $i$ and $j$ played $L$, for particular $1 \leq i < j \leq 4$. Therefore, $\operatorname{conv}\mathfrak{B}^{\mathrm{S}}$ includes all possible beliefs at $U$. $\mathfrak{B}^{\mathrm{F}}$ differs from both sets, since (i) it includes the uniform distribution on the nodes in $U$, and (ii) all its elements satisfy $p_{12}p_{34} = p_{13}p_{24} = p_{14}p_{23}$, because similar equalities hold under every completely mixed strategy profile.

probability $x^{\epsilon}(a)$ to each action $a$:

$$x^{\epsilon}(a) = \left(1 - \epsilon - \epsilon^{|\mathcal{A}|+1}\right)x(a) + \epsilon x^{\mathcal{A}}(a) + \epsilon^{|\mathcal{A}|+1}z(a), \tag{6}$$

where $x(a)$, $x^{\mathcal{A}}(a)$ and $z(a)$ are the probabilities specified for $a$ by $x$, $x^{\mathcal{A}}$ and $z$ respectively. In this context, possible outcomes of chance moves are also viewed as "actions," whose (positive, by assumption) probabilities are fixed and are the same in all strategy profiles. The unique belief system $\mu^{\epsilon}$ that is weakly consistent with the completely mixed strategy profile $x^{\epsilon}$ satisfies

$$\mu^{\epsilon}(V) = \mathbb{P}_{x^{\epsilon}}(V \mid U) = \frac{\mathbb{P}_{x^{\epsilon}}(V)}{\mathbb{P}_{x^{\epsilon}}(U)}, \qquad V \subseteq U. \tag{7}$$

For every node $v \in U$, $\mathbb{P}_{x^{\epsilon}}(\{v\}) = \prod_l x^{\epsilon}(a^l)$, where the $a^l$'s are all the actions preceding node $v$. In view of (6), this product can be expressed as a polynomial in $\epsilon$, $\sum_{j \geq 0} \beta_j \epsilon^j$. For $j < |\mathcal{A}|$, the coefficient $\beta_j$ is zero, because a positive coefficient would mean that $\mathbb{P}_{x^{\mathcal{B}}}(\{v\}) > 0$ for some $\mathcal{B} \subsetneq \mathcal{A}$, which contradicts the minimal-sufficiency assumption concerning $\mathcal{A}$. By a similar argument, $\beta_{|\mathcal{A}|} = \mathbb{P}_{x^{\mathcal{A}}}(\{v\})$. It follows that, for $V \subseteq U$, $(1/\epsilon^{|\mathcal{A}|})\,\mathbb{P}_{x^{\epsilon}}(V) \to \mathbb{P}_{x^{\mathcal{A}}}(V)$ as $\epsilon \to 0$, which implies that the quotient in (7) converges to $\mathbb{P}_{x^{\mathcal{A}}}(V) \,/\, \mathbb{P}_{x^{\mathcal{A}}}(U)$. Therefore, if $(\epsilon_k)_{k=1}^{\infty}$ is any sequence of positive numbers converging to 0 such that $(\mu^{\epsilon_k})_{k=1}^{\infty}$ converges to some limit $\mu$, then the belief system $\mu$, which is clearly fully consistent with $x$, satisfies (4). The existence of such a sequence follows from the obvious compactness of the set of all belief systems.

To prove (iv), consider any belief system $\mu$ that is fully consistent with $x$ and some sequence $(\mu^k, x^k)_{k=1}^{\infty}$ as in the definition of full consistency. For each $\mathcal{A} \in \mathfrak{A}$, let $u \in U$ be as in (i). For every $k$, $\mathbb{P}_{x^k}(\{u\}) = \prod_l x^k(a^l)$, where the $a^l$'s are all the actions (and outcomes of chance moves) preceding $u$, and so

$$\frac{1}{\prod_{a \in \mathcal{A}} x^k(a)}\,\mathbb{P}_{x^k}(\{u\}) = \prod_{\substack{l \\ a^l \notin \mathcal{A}}} x^k(a^l) \underset{k \to \infty}{\longrightarrow} \mathbb{P}_{x^{\mathcal{A}}}(\{u\}). \tag{8}$$

A result similar to (8) holds with $u$ replaced by any other node $v \in U$ that is preceded by all the actions in $\mathcal{A}$. Therefore, for such $v$,

$$\frac{\mathbb{P}_{x^{\mathcal{A}}}(\{v\} \mid U)}{\mathbb{P}_{x^{\mathcal{A}}}(\{u\} \mid U)} = \frac{\mathbb{P}_{x^{\mathcal{A}}}(\{v\})}{\mathbb{P}_{x^{\mathcal{A}}}(\{u\})} = \lim_{k \to \infty} \frac{\mathbb{P}_{x^k}(\{v\})}{\mathbb{P}_{x^k}(\{u\})} = \lim_{k \to \infty} \frac{\mu^k(\{v\})}{\mu^k(\{u\})} = \frac{\mu(\{v\})}{\mu(\{u\})}$$

if $\mu(\{u\}) > 0$, and if $\mu(\{u\}) = 0$, then $\mu(\{v\}) = 0$ (for otherwise the second limit above would not exist). It follows that $\mu(\{v\}) = \lambda_{\mathcal{A}}\,\mathbb{P}_{x^{\mathcal{A}}}(\{v\} \mid U)$, where $\lambda_{\mathcal{A}} \coloneqq \mu(\{u\})/\mathbb{P}_{x^{\mathcal{A}}}(\{u\} \mid U)$. For all $\mathcal{A}' \neq \mathcal{A}$ in $\mathfrak{A}$, $\mathbb{P}_{x^{\mathcal{A}'}}(\{v\} \mid U) = 0$ (because $x^{\mathcal{A}'}$ assigns probability zero to the actions in $\mathcal{A} \setminus \mathcal{A}'$), and so the equality in (5) holds for $V = \{v\}$.

To prove that the equality holds generally, it remains to note that every $v \in U$ is preceded by all the actions in *some* element of $\mathfrak{A}$, because the set of actions that precede $v$ and are assigned probability zero by $x$ necessarily has a subset that is minimally sufficient for reaching $U$ under $x$.

Setting $V = U$ in (5) proves that the coefficients sum up to 1. ∎

## 2.1 Structural Consistency

Reaching an off-path information set $U$ may also be explainable by deviations from the players' strategy profile $x$ that involve a non-minimally sufficient set of actions. However, such an explanation represents a non-parsimonious hypothesis as to why $U$ was reached; it assumes more than it has to. Moreover, the explanation may have the troubling aspect that it implies a *future* deviation from $x$. This can happen if some players' information sets include both nodes that precede $U$ and nodes that follow it.

**Example 1** (Kreps and Ramey 1987) In the game in Figure 1c, the players' order of moves is uncertain — the player moving before player 2 is either 1 or 3 — and is unknown to them. It is not difficult to see that the only Nash equilibrium outcome is that players 1 and 3 "quit" by playing $Q'$ and $Q''$ for sure, so that player 2's information set is not reached. The choice to quit reflects the fact that, in every Nash equilibrium, the probabilities that player 2's strategy assigns to playing $L$ and $R$ are not greater than $1/3$. In particular, neither probability is 1. Such a strategy can be justified only by beliefs that attach positive probability to both nodes in player 2's information set, and such beliefs are induced only by strategy profiles in which both player 1 and player 3 deviate from their equilibrium strategies by "proceeding" (playing $P'$ and $P''$) with positive probability. However, such simultaneous deviations are inconsistent with an assumption that the player acting *after* player 2 will be using his equilibrium strategy.

Example 1 shows that a non-parsimonious hypothesis about the past may project onto the future. With a parsimonious hypothesis about the deviations that led to an information set being reached, this cannot happen.

**Proposition 1** If a set of actions $\mathcal{A}$ is minimally sufficient for reaching an information set $U$ of a player $i$ under a strategy profile $x$, then (i) $U$ is reached with positive probability under $x^{\mathcal{A}}$, and (ii) $x$ and $x^{\mathcal{A}}$ agree at $U$ and at every other information set $U'$ such that, under $x^{\mathcal{A}}$ or some other strategy profile that differs from $x^{\mathcal{A}}$ only in player $i$'s strategy, $U'$ is reached with positive probability after $U$ is reached.

*Proof.* Assertion (i) holds by definition. To prove (ii), consider any strategy profile that differs from $x^{\mathcal{A}}$, if at all, only in the strategy of player $i$. Any path that has positive probability under that strategy profile and reaches $U$ must, by the minimal-sufficiency assumption, first go through *all* the actions in $\mathcal{A}$. By the perfect-recall assumption, the path cannot revisit any of the information sets where these actions were taken. ■

Kreps and Ramey (1987) call a belief system $\mu$ *structurally consistent*[5] with a strategy profile $x$ if the beliefs specified by $\mu$ at every information set $U$ satisfy augmented weak consistency and coincide with $\mathbb{P}_{x'}(\cdot \mid U)$ (i.e., with the beliefs obtained using Bayes' rule) for some strategy profile $x'$ (which may depend on $U$) that has the two properties (i) and (ii) specified for $x^{\mathcal{A}}$ in Proposition 1. The proposition thus shows that strongly consistent beliefs satisfy structural consistency.

[5] Note that theirs is a stronger concept than Kreps and Wilson's (1982) earlier notion of structural consistency of beliefs, for which the strategy profile is irrelevant.

A belief system $\mu$ is *convex structurally consistent* with a strategy profile $x$ if there is a finite set of structurally consistent belief systems such that the beliefs specified by $\mu$ at every information set $U$ are a convex combination of those specified by these belief systems (with weights that may depend on $U$). The Proposition in Kreps and Ramey (1987) asserts that every fully consistent assessment (hence every sequential equilibrium) satisfies convex structural consistency. Proposition 1 and part (iv) of Lemma 1 ($\mathfrak{B}^{\mathrm{F}} \subseteq \mathrm{conv}\ \mathfrak{B}^{\mathrm{S}}$) strengthen this result. They show that, for every fully consistent assessment, the beliefs at every information set are in fact a convex combination of the strongly consistent beliefs.

The beliefs that justify any sequential (or perfect Bayesian) equilibrium strategy for player 2 in Example 1 are only convex structurally consistent. They can be obtained only as convex combinations of the two structurally (and strongly) consistent beliefs, which are those that assign probability 1 to one of player 2's nodes and reflect a hypothesized deviation by only one, particular other player. The example thus demonstrates the unavoidability of dealing with structurally inconsistent beliefs in sequential and perfect Bayesian equilibria. Agnostic sequential equilibrium, by contrast, always involves structurally consistent beliefs at every information set $U$. The significance of this fact is that, as shown, such beliefs about past actions do not "taint" the continuation game starting at $U$. That is, they never imply that some players' strategies in that game are necessarily different from those induced by the original strategies in the whole game.

# 3 Agnostic Sequential Equilibrium

The discussion in the previous sections leads to the following formal definition, where strongly consistent beliefs are expressed as appropriately modified strategy profiles (see Definition 1). The payoff function of a player $i$ is denoted $u_i$.

**Definition 2** A strategy profile $x$ is an *agnostic sequential equilibrium (ASE)* if for every player $i$, information set $U$ of that player and action $a$ in $U$,

$$u_i(x^{\mathcal{A}}) \geq u_i(x^{\mathcal{A} \cup \{a\}}) \tag{9}$$

for every set of actions $\mathcal{A}$ that is minimally sufficient for reaching $U$ under $x$.

The requirement in Definition 2 means that, under any strongly consistent beliefs at any information set $U$, the player $i$ acting at $U$ cannot increase his payoff in the continuation game by choosing a different action at $U$ than that specified by his strategy $x_i$. This may seem to leave open the possibility that the payoff can be increased by also changing the action at some later information set(s). However, the next theorem shows that Definition 2 actually excludes this possibility. [6]

[6] The characterization in Theorem 1 could alternatively be taken as the definition of ASE. The equivalence with Definition 2 means that ASE, like sequential equilibrium and similar solution concepts, has the *one-deviation property* (Hendon et al. 1996; Perea 2002).

**Notation** For a strategy profile $x$, a player $i$, an information set $U$ of that player and a strategy $x_i'$, the strategy profile that differs from $x$ only in that, at $U$ and all the information sets that follow it, player $i$ plays according to $x_i'$ is denoted $x \mid_U x_i'$.

**Theorem 1** A strategy profile $x$ is an agnostic sequential equilibrium if and only if for every player $i$, information set $U$ of that player and strategy $x_i'$,

$$u_i(x^{\mathcal{A}}) \geq u_i(x^{\mathcal{A}} \mid_U x_i') \tag{10}$$

for every set of actions $\mathcal{A}$ that is minimally sufficient for reaching $U$ under $x$.

*Proof.* The sufficiency of the condition is obvious, as (9) is a special case of (10), with $x_i'$ that differs from $x_i$ only in specifying that $i$ takes the action $a$ at $U$. To prove necessity, suppose that the condition does not hold: for some $i$, $U$, $x_i'$ and $\mathcal{A}$ as above, $u_i(x^{\mathcal{A}} \mid_U x_i') > u_i(x^{\mathcal{A}})$. Without loss of generality, it can be assumed that $x_i'$ is a pure strategy and so the left-hand side of the last inequality can be written as $u_i(x^{\mathcal{A}\cup\mathcal{B}})$, where $\mathcal{B}$ is a set whose elements are actions of player $i$ at $U$ or later information sets. Also without loss of generality, $\mathcal{B}$ is minimal, that is, $u_i(x^{\mathcal{A}\cup\mathcal{B}'}) \leq u_i(x^{\mathcal{A}})$ for all $\mathcal{B}' \subsetneq \mathcal{B}$. In particular, for every $a \in \mathcal{B}$ that is playable at an information set $U'$ that does not precede any other information set where an element of $\mathcal{B}$ is playable, $u_i(x^{\mathcal{A}\cup\mathcal{B}'}) \leq u_i(x^{\mathcal{A}}) < u_i(x^{\mathcal{A}\cup\mathcal{B}})$ holds for $\mathcal{B}' = \mathcal{B} \setminus \{a\}$. These two inequalities imply that $U'$ is reached with positive probability under $x^{\mathcal{A}\cup\mathcal{B}'}$ and that deviating by playing $a$ at $U'$ increases player $i$'s payoff in the continuation game (with the beliefs at $U'$ given by Bayes' rule). The same is true with $\mathcal{A} \cup \mathcal{B}'$ replaced with any set $\mathcal{A}'$ that is minimally sufficient for reaching $U'$ under $x$ and satisfies $\mathcal{A} \subseteq \mathcal{A}' \subseteq \mathcal{A} \cup \mathcal{B}'$ (such a set necessarily exists, because $U'$ either coincides with or follows $U$ and therefore any subset of $\mathcal{A} \cup \mathcal{B}'$ that is minimally sufficient for reaching $U'$ must contain $\mathcal{A}$), as changing from $x^{\mathcal{A}\cup\mathcal{B}'}$ to $x^{\mathcal{A}'}$ does not change the beliefs at $U'$ (see (2)). Thus, $u_i(x^{\mathcal{A}'}) < u_i(x^{\mathcal{A}'\cup\{a\}})$, which shows that $x$ is not an ASE. ∎

## 3.1 ASE and Sequential Equilibrium

If the notion of strong consistency of beliefs and a strategy profile implicit in Definition 2 were replaced with an even stronger, less inclusive kind of consistency, the result would be a *weaker* definition of agnostic sequential equilibrium. That is, the set of qualifying strategy profiles would expand or remain unchanged. The opposite is true for any weaker notion, such as weak or augmented weak consistency. As indicated, the notion of local consistency corresponding to full consistency is also weaker than strong consistency (that is, at every information set, $\mathfrak{B}^{\mathrm{S}} \subseteq \mathfrak{B}^{\mathrm{F}}$). However, replacing the latter with the former would actually not change the meaning of ASE. This conclusion follows from the second inclusion in (3), $\mathfrak{B}^{\mathrm{F}} \subseteq \mathrm{conv}\ \mathfrak{B}^{\mathrm{S}}$, because if each element in a particular set of beliefs at an information set $U$ justifies the choice (or the exclusion) of a particular strategy in the continuation game, then any convex combination of these beliefs automatically justifies it too. The conclusion shows that, fundamentally, the difference between sequential equilibrium and ASE does not stem from the different notions of consistency they employ but is wholly due to the logical difference between requiring sequential rationality with respect to *some* consistent beliefs and *all* such beliefs, respectively.

**Theorem 2** A strategy profile $x$ is an agnostic sequential equilibrium if and only if the assessment $(\mu, x)$ is a sequential equilibrium for *every* belief system $\mu$ that is fully consistent with $x$.

*Proof.* Inequality (10) requires player $i$'s expected payoff in the game to be the same or higher under $x^{\mathcal{A}}$ than under $x^{\mathcal{A}} \mid_U x_i'$. Clearly, an equivalent condition is that a similar inequality holds for the *conditional* expectation of the payoff, given that the information set $U$ was reached. Both conditional expectations are the expected payoffs in the continuation game starting at $U$, with the (strongly consistent) beliefs there induced under $x$ by the minimally sufficient set of actions $\mathcal{A}$. In view of Proposition 1, the first expected payoff is obtained when the strategy profile in the continuation game is (that induced by) $x$ and the second one is obtained with $x \mid x_i'$. It therefore follows from (iii) in Lemma 1 ($\mathfrak{B}^{\mathrm{S}} \subseteq \mathfrak{B}^{\mathrm{F}}$) that if $(\mu, x)$ is a sequential equilibrium for every belief system $\mu$ that is fully consistent with $x$, then the condition in Theorem 1 holds. It follows from (iv) in that lemma ($\mathfrak{B}^{\mathrm{F}} \subseteq \mathrm{conv}\ \mathfrak{B}^{\mathrm{S}}$) that, conversely, if the latter condition holds, then the former holds. ■

**Example 2** As indicated, for the strategy profiles shown in Figure 1a and b, both $p = 0$ and $p = 1$ are strongly, hence fully, consistent beliefs. Together with the former belief ($p = 0$), these strategy profiles constitute sequential equilibria, but this is not so for the latter. Therefore, the two strategy profiles are not agnostic sequential equilibria. Moreover, no ASE exists in the game in Figure 1a if $\alpha > 2$.

The two sequential equilibria in Example 2 are furthermore perfect equilibria (Selten 1975) and quasi-perfect equilibria (Van Damme 1984).[7] The example therefore shows that these solution concepts, which are both stronger than sequential equilibrium, do not imply agnostic sequential equilibrium.[8] The reverse implications also do not hold. In particular, it follows immediately from the definition that any Nash equilibrium where all information sets are reached with positive probability is an ASE, but such an equilibrium is not necessarily perfect or quasi-perfect. For example, with $\alpha = 2$ in the game in Figure 1a, the strategy profile $(L, l)$ is an ASE but it is not a perfect or quasi-perfect equilibrium. The underlying reason is that both flavors of perfectness are based on the idea that players *expect* other players to make mistakes (albeit with very small probability), which in particular entails that a weakly dominated strategy such as $L$ should not be chosen. In sequential equilibrium and ASE, by contrast, possible mistakes are considered only at off-equilibrium information sets, where it is evident that a deviation has actually occurred.

[7] They are moreover proper equilibria (Myerson 1978) in both the normal form and the agent normal form of the respective games. To see that the strategy profile $(L, QQ', r)$ is a proper equilibrium in the normal form of the game in Figure 1b, note that it is the limit as $\epsilon \to 0$ of the strategy profile where $R$ is chosen with probability $2\epsilon$, strategies $QP'$, $PQ'$ and $PP'$ are chosen with probability $\epsilon$, $\epsilon^2$ and $\epsilon^3$, respectively, and $l$ is chosen with probability $\epsilon$. For $0 < \epsilon < 1/6$, $L$ is a dominant strategy, the strategies $QQ'$, $QP'$, $PQ'$ and $PP'$ yield progressively lower payoffs, and $r$ yields $2\epsilon^3 + 6\epsilon^4$ more than $l$ does. It follows that the probability assigned to every strategy is less than $3\epsilon$ times the probability assigned to any better strategy, which by definition means that the limit is a proper equilibrium.

[8] This conclusion also follows from the fact that a proper equilibrium exists in every normal form game.

# 4 Beyond Finite Trees

As stated in Section 1, the formal setting of the above discussion is that of extensive-form games. However, the actual definition of agnostic sequential equilibrium is also applicable to dynamic games that cannot be described by a finite game tree, in particular, games with a continuum of actions.[9] In this respect, ASE is more similar to perfect Bayesian equilibrium than to sequential equilibrium.

Definition 2 is formally meaningful even if minimally sufficient sets are not guaranteed to exist, which can happen with a continuum of outcomes to chance moves. However, in games with a continuum of actions, agnostic sequential rationality may run into conceptual problems similar to those plaguing more traditional forms of sequential rationality, as described by Myerson and Reny (2015, 2020). These problems stem from the fact that reaching a zero-probability information set may actually be expected and does not necessarily indicate a deviation.

**Example 3** (Myerson and Reny 2015) Player 1 can play $L$ or $R$. If he chooses $L$, then there is a chance move in which a number is drawn uniformly from the unit interval $[0,1]$. If he chooses $R$, then the player himself has to choose a number in $[0,1]$. Player 2 is told the number $s$ selected, but not whether it resulted from a chance move or was chosen by player 1. His possible actions are also $L$ and $R$, and the payoffs are as in the battle of the sexes game. In particular, only $(L, L)$ and $(R, R)$ yield a positive payoff to either player. In BoS, both strategy profiles are equilibria. Here, however, $(R, R)$ is played with probability 1 in any agnostic sequential equilibrium. To see this, consider the information set of player 2 where he is informed that a particular $s \in [0,1]$ was selected. The probability that a chance move yields $s$ is zero. Therefore, regardless of player 1's actual strategy (and whether or not it involves randomization), there is a unique set of actions that is minimally sufficient for reaching player 2's information set, which corresponds to the belief that player 1 played $R$ and then chose $s$. Player 2 should therefore play $R$ regardless of $s$, and player 1's best response to this is also playing $R$. Thus, the strategy profile where player 1 plays $L$ and player 2 also plays $L$ regardless of $s$ is not an ASE, even though the actions best respond to one another and, intuitively, because all outcomes of the change move are equally likely, no value of $s$ should raise a suspicion of deviation by player 1.

One wonders whether the problem illustrated by Example 3 is but a modeling artifact. An implicit premise here is that it is possible for a player or a randomization device to select an arbitrary real number and transmit it to the relevant parties. However, it may be argued that this task is feasible only if the selected number has a "name". This is so for any integer, any algebraic (in particular, rational) number and any number that can be generated using the algebraic and the few named transcendental numbers, such as $\pi$ or $e$. An arbitrary real number can be specified and transmitted bit-by-bit, using its binary expansion. However, this requires infinite time, and so there are only finitely many possible outcomes in any finite time period. In principle, randomization that results in an arbitrary real number can be performed using a physical device such as a roulette wheel. However, reading the outcome can be performed only with finite precision, and while this limitation may be of little practical concern, it is another conceptual problem with a continuum of possible outcomes.

---

[9] Definition 2 is, in addition, meaningful without perfect recall. That property, however, is assumed in Theorem 1.

The problem with generating and transmitting the outcome does not arise when the choice is from a closed list, either finite or countably infinite. A continuum of outcomes may represent the limiting case in which the resolution provided by the list approaches infinity. However, a continuum is hardly an ideal approximation if it gives rise to difficulties that are not shared by the finite or countably infinite cases. The difficulty here is of this nature. With at most countably infinitely many possible outcomes, each with a specified probability, a zero-probability outcome may be expected *not* to be obtained, and so the observation of such an outcome can and should be interpreted as indicating a deviation. For the continuum, the same is not true.

# 5 Beliefs Based on Strategic Reasoning

Strong consistency is based on the principle of parsimony: a particular deviation from the players' strategies is assumed to have occurred only if this assumption is needed for explaining the arrival at an off-path information set. However, the simplest explanation is not always the most convincing one. In particular, *forward induction* arguments (Kohlberg and Mertens 1986) may lend credence to inconsistent beliefs. That is, a detected past deviation of another player from his strategy may hint at an additional, unobservable deviation. Unlike strong consistency, which is a notion based wholly on the form of the game tree, forward induction also involves examination of the strategic interests of the deviating players. Such an examination may suggest an intentional deviation (Van Damme 1991, Section 10.5). For example, in Figure 2a, player 2's choice of $r$ in the agnostic sequential equilibrium Black is supported by the unique strongly (and unique fully) consistent belief, which is that player 1 would follow his strategy in the proper subgame and choose $R$ there. However, if 2's information set is actually reached, which indicates that player 1 deviated from his strategy in the whole game by playing $In$, player 2 may reason that the most likely explanation for the deviation is that player 1 is aiming for the better equilibrium Gray, and thus also deviated by playing $L$ rather than $R$. Such a belief makes $l$ the better choice for player 2.

Past deviations may also be taken as indicators of intended *future* ones. This possibility is illustrated by the game the differs from that in Figure 2a only in the order of moves in the subgame: player 2 chooses his action before player 1 does so. As the two moves are actually effectively simultaneous, this game is essentially identical to the one considered above, and so the same argument applies: if player 1 deviated once, strategic considerations suggest he intends to deviate again.

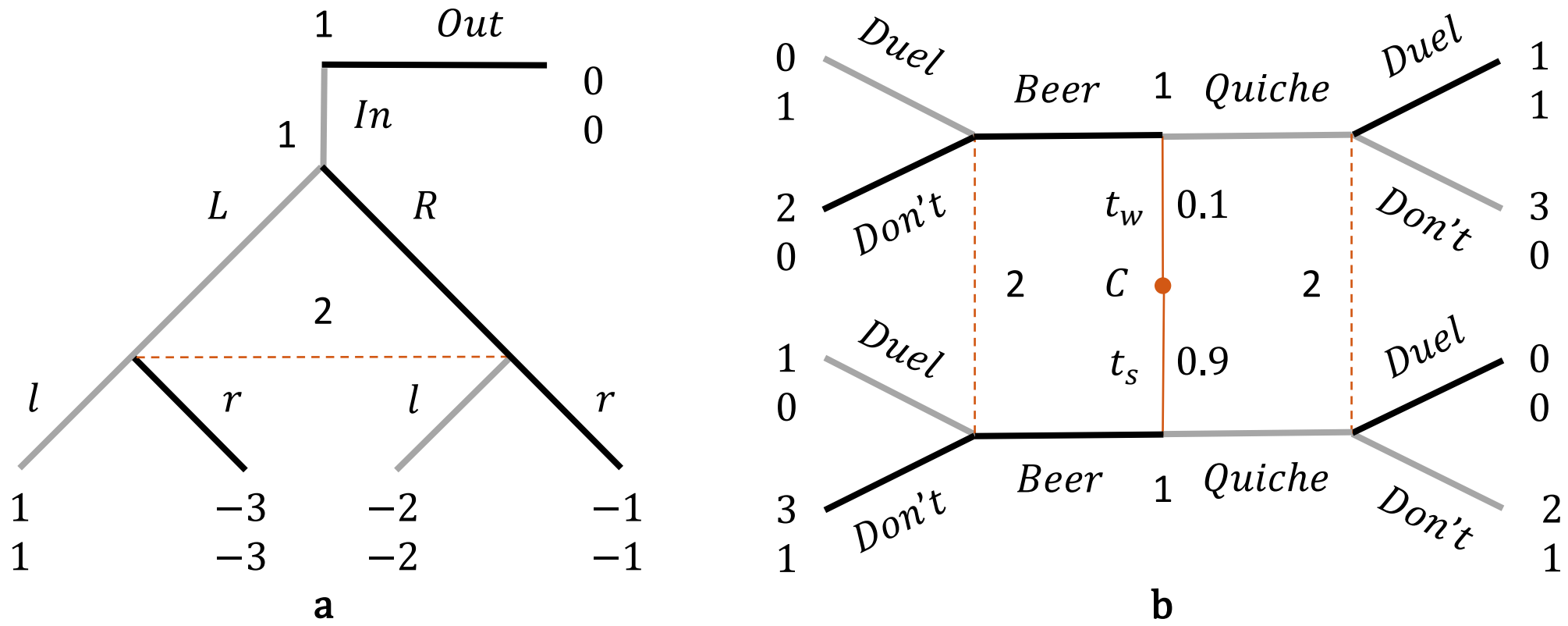

**FIGURE 2 THE DESTRUCTIVE AND CONSTRUCTIVE POTENTIAL OF BELIEFS REFLECTING STRATEGIC REASONING. a BOTH BLACK AND GRAY (INDICATED BY LINES OF THAT COLOR) ARE AGNOSTIC SEQUENTIAL EQUILIBRIA, IN WHICH PLAYER 1'S SPECIFIED ACTION IN THE PROPER SUBGAME INDUCES UNIQUE STRONGLY CONSISTENT BELIEFS FOR PLAYER 2. HOWEVER, BLACK IS PUT INTO QUESTION BY THE FACT THAT A DEVIATION BY PLAYER 1 THAT LEADS TO THE SUBGAME BEING REACHED MAY SUGGEST AN ADDITIONAL DEVIATION THERE: AN ATTEMPT TO GET A POSITIVE PAYOFF BY PLAYING $L$. b THE BEER-QUICHE GAME. BLACK AND GRAY ARE SEQUENTIAL EQUILIBRIA, BUT ONLY THE FORMER SATISFIES THE INTUITIVE CRITERION. BOTH EQUILIBRIA ARE NOT AGNOSTIC SEQUENTIAL EQUILIBRIA. HOWEVER, ADOPTING THE RESTRICTION ON OFF-EQUILIBRIUM BELIEFS UNDERLYING THE INTUITIVE CRITERION WOULD MAKE BLACK AN ASE.**

A reasonable general criterion that takes the players' incentives into consideration is that an explanation for reaching an off-equilibrium information set which does not involve some player taking a strictly dominated action or strategy should be favored over an explanation that does so. In games with multiple sequential or perfect Bayesian equilibria, a restriction of this kind on off-equilibrium beliefs may eliminate some of the equilibria. For example, an extension of the idea that beliefs at an information set should assign positive probability only to those nodes that are reached with the smallest number of strictly dominated actions leads to the notions of *justifiable beliefs* and *justifiable equilibrium*, the latter defined as any sequential equilibrium with justifiable beliefs (McLennan 1985). As pointed out by McLennan, Black in Figure 2a actually *is* a justifiable equilibrium, as none of the actions in the game is dominated. Nevertheless, adopting the above general criterion does eliminate this agnostic sequential equilibrium, because it renders player 2's belief that player 1 played the strictly dominated strategy $In\ R$ unreasonable. Although the alternative belief that 1 played $In\ L$ is not strongly consistent with the player's equilibrium strategy $Out\ R$, it is deemed more compelling because $In\ L$ is undominated.

A different potential outcome of strategic reasoning is selection among beliefs that *are* consistent with the players' strategies. Consider, for example, the Spence education model with two types of worker, where a restriction on off-equilibrium beliefs can eliminate all pooling perfect Bayesian equilibria (Cho and Kreps 1987; Mas-Colell et al. 1995). For agnostic sequential equilibrium, restrictions on off-equilibrium beliefs may be particularly called for in this model, because none of the pooling equilibria is an ASE to begin with. The reason is that the choice of any off-equilibrium education level cannot be excluded, as the requirement of strong consistency does not preclude a belief by an employer that such a choice indicates a high quality worker.[10] However, the effect of selection among beliefs on the ASE solution concept is in a sense the diametric opposite of the effect on sequential and perfect Bayesian equilibria. As any such selection effectively entails a stronger notion of consistency than strong consistency, it cannot eliminate ASEs but can only *add* new ones (see the remark at the beginning of Section 3.1). Thus, the set of agnostic sequential equilibria, which is typically contained in the sets of perfect Bayesian and sequential equilibria (see Theorem 2), expands while the latter contract, which means that the gap between the corresponding sets of results may narrow. The following example illustrates this possibility.

**Example 4** Consider the Beer-Quiche game shown in Figure 2b, where for simplicity only pure strategies are allowed. There are two Nash equilibria, Black and Gray, which are both pooling sequential equilibria. In Black, types $t_w$ and $t_s$ of player 1 both choose $Beer$, and in Gray, they choose $Quiche$. The second equilibrium is eliminated by the *intuitive criterion* (Cho and Kreps 1987). The criterion is based on a restriction of player 2's possible beliefs regarding player 1's type, which in particular precludes beliefs that, following a choice of $Beer$, attach a positive probability to $t_w$. The reason such beliefs are deemed unreasonable is that this type's equilibrium payoff of 3 is higher than anything he may get by choosing $Beer$. The same problem does not arise in Black, where both types of player 1 choose $Beer$ and player 2 would choose $Duel$ only as a response to $Quiche$. This response is justified by the unique reasonable belief following a choice of $Quiche$ by player 1, which is that his type is $t_w$ (because $t_s$ would necessarily be harmed by such a choice). The same argument also shows that a restriction to reasonable beliefs would make Black an agnostic sequential equilibrium, as the strongly consistent off-equilibrium belief that it was $t_s$ who chose $Quiche$ is discarded.

Thus, the logic underlying the intuitive criterion singles out the same equilibrium for both solution concepts, sequential and agnostic sequential equilibrium. This coincidence contrasts with the situation for the original, unmodified definitions, according to which both equilibria are sequential equilibria but neither of them is an ASE. It is, however, a rather special outcome, which is due to the fact that the additional reasonableness requirement on off-equilibrium beliefs pins them down uniquely.

# 6 Agnostic Sequential Polyequilibrium

In an off-path information set, where there may be multiple strongly consistent beliefs, there may be no action that is optimal under all these beliefs. This fact raises the possibility that the action at such an information set may have to be left at least partially unspecified. Doing so leads to a set-valued solution concept. Instead of a single strategy profile $x$, the solution is a set of strategy profiles $X$.

[10] More generally, in an agnostic sequential equilibrium, a player would not know what to make of a signal that the sender's strategy never specifies; such a signal may mean anything. In this, ASE differs from perfect Bayesian equilibrium, which requires a ready interpretation for every physically possible signal.

An example of such a solution concept is *essentially perfect Bayesian equilibrium* (Blume and Heidhues 2006). In an incomplete-information game with perfect recall, an EPBE is a set $X$ that is obtained from a specified strategy profile $x$ by declaring certain information sets *relevant* and the remaining ones *irrelevant*. The collection of irrelevant information sets must (i) have zero probability of being reached when $x$ is played, and (ii) include every information set that follows any irrelevant information set of the same player. The set $X$ is then defined as the collection of all strategy profiles that agree with $x$ at each of the relevant information sets. By requirement (i), any belief system $\mu$ that is weakly consistent with $x$ is also weakly consistent with every other element of $X$. The definition of EPBE is completed by the requirement that there is some weakly consistent belief system $\mu$ as above (which may be considered part of the EPBE) such that, for every relevant information set $U$ of every player $i$, strategy $x_i$ is a best response to *all* strategy profiles in $X$ in the continuation game starting at $U$ with the beliefs specified there by $\mu$.

Essentially perfect Bayesian equilibrium extends perfect Bayesian equilibrium in being a set-valued solution concept. However, it is still based on a single, possibly arbitrary, belief system. Thus, it is not in the spirit of, and it does not extend, agnostic sequential equilibrium. To extend the latter, a more general set-valued solution concept is needed. Polyequilibrium (Milchtaich 2019) is such a concept.[11]

In a simultaneous-move (i.e., normal form) or dynamic game, where each player $i$ has a strategy set $S_i$ and a payoff function $u_i$, a *polystrategy* for player $i$ is any nonempty set of strategies $\emptyset \neq X_i \subseteq S_i$. A *polystrategy profile* $X$ is a Cartesian product of polystrategies, one polystrategy $X_i$ for each player $i$. In other words, it is a nonempty rectangular set of strategy profiles. $X$ is a *polyequilibrium* if for every player $i$ and (excluded) strategy $x_i' \notin X_i$ there is some (non-excluded) $x_i'' \in X_i$ that responds to all strategy profiles in $X$ at least as well as $x_i'$ does, that is,

$$u_i(x \mid x_i'') \geq u_i(x \mid x_i'), \qquad x \in X.$$

Polyequilibrium is an "excluding" solution concept. It requires justification for the exclusion of the strategies that are not in a player's polystrategy rather than for the inclusion of those in it. This characteristic of polyequilibrium bears on its informational content. The content is conveyed by the concept of polyequilibrium result (Milchtaich 2019).

In any game, a *result* $R$ is any set of strategy profiles. The result *holds* in a polystrategy profile $X$ if $X \subseteq R$, and it is a *polyequilibrium result* if it holds in some polyequilibrium in the game. A result may also be specified implicitly, as a particular property or consequence of strategy profiles (for example, "player 1's payoff is higher than 2's payoff"). In this case, $R$ is the collection of all strategy profiles that have the specified property, so that the result holds in a polyequilibrium $X$ if and only if *all* strategy profiles in $X$ have the property. In particular, a real number $v_i$ is a *polyequilibrium payoff* for a player $i$ if there is some polyequilibrium $X$ with $u_i(x) = v_i$ for all $x \in X$, and a strategy $x_i$ is a *polyequilibrium strategy* if there is some polyequilibrium $X$ with $X_i = \{x_i\}$.

[11] EPBE is a special kind of polyequilibrium, indeed of *simple* polyequilibrium. A simple polyequilibrium always includes at least one Nash equilibrium, whereas in a general polyequilibrium it is possible that none of the elements is an equilibrium. This possibility makes polyequilibrium substantially different from set-valued solution concepts like *strategic stability* (Kohlberg and Mertens 1986) where the set includes *only* equilibria.

In a dynamic context, the polyequilibrium condition may be naturally strengthened by the addition of the agnostic sequential rationality requirement. That is, for every excluded strategy $x_i'$ there must be an non-excluded strategy $x_i''$ that is an adequate substitute for $x_i'$ in the continuation game starting at each of player $i$'s information sets $U$, for every strategy profile $x \in X$ and every probability distribution over the nodes in $U$ that is strongly consistent with $x$. Beliefs that are not strongly consistent are thus excluded. As for strategies, the non-excluded, consistent beliefs may or may not be justifiable (see Section 5). These ideas lead to the following definition.

**Definition 3** A polystrategy profile $X$ is an *agnostic sequential polyequilibrium (ASPE)* if for every player $i$ and strategy $x_i' \notin X_i$ there is some $x_i'' \in X_i$ such that, for every information set $U$ of player $i$,

$$u_i(x^{\mathcal{A}} \mid_U x_i'') \geq u_i(x^{\mathcal{A}} \mid_U x_i') \tag{11}$$

for every strategy profile $x \in X$ and every set of actions $\mathcal{A}$ that is minimally sufficient for reaching $U$ under $x$.

Agnostic sequential polyequilibrium is both a refinement of polyequilibrium (in the context of dynamic games) and a generalization of agnostic sequential equilibrium. A singleton $\{x\}$ is an ASPE if and only if the strategy profile $x$ is an ASE.

Of the two solution concepts, ASPE may be viewed as the principal one and ASE as a mere appendage. For the former, unlike the latter, existence is not an issue. Indeed, the set of all strategy profiles is an agnostic sequential polyequilibrium, called the *trivial* ASPE. With respect to the inclusion order, the trivial ASPE is the largest element while the ASEs (being singletons) are among the minimal ones. Restricting attention to minimal ASPEs may seem like a natural, standard approach. However, as argued in Milchtaich (2019), doing so would actually be counter-productive. The indeterminacy allowed by the polyequilibrium concept is an asset, not a liability — especially in the dynamic-game context.

A simple example of the informational content of ASPE is provided by the game in Figure 1a with $\alpha > 2$. As indicated, this game has no ASE. However, it has an obvious non-trivial ASPE, which is the one presented in Section 1. In this polyequilibrium, the strategy of player 2 is not specified but player 1 plays $Out$, which is therefore an agnostic sequential polyequilibrium strategy.

Another example is the following one.

**Example 5** There is a \$1 bill in either the blue or the yellow box, and the two cases, $B$ and $Y$, are equally likely. The actual location of the money is private information of player 1, but only player 2 can to open a (single) box and take the money if it is there. Player 1 has to choose the price $p \geq 0$ at which he offers to sell the information to player 2. The latter can then pay the price and open the correct box, reject the offer and open the blue box, reject it and open the yellow box, or reject and flip a fair coin to choose a box. For simplicity, (other) mixed strategies are not allowed.

Every sum $0 \leq v_1 \leq 1/2$ is a perfect Bayesian equilibrium payoff for player 1. It is obtained in an equilibrium where player 1's price is $v_1$ and player 2 is willing to pay this price but would reject the offer and open the blue box if player 1 asked a different price. This reaction is supported by a belief that a price different from $v_1$ indicates that the money is in the blue box.

By contrast, the only agnostic sequential polyequilibrium payoff for player 1 is $1/2$. To see this, suppose that there is an ASPE $X$ where the payoff is $v_1 < 1/2$, and consider some strategy $x_1 \in X_1$ and some price $v_1 < p < 1/2$ that is different from the two prices that $x_1$ specifies for the two cases $B$ and $Y$ (which in principle can be different). Player 2's polystrategy $X_2$ necessarily excludes acceptance of price $p$, for if it included a strategy prescribing acceptance, it would not be possible to exclude player 1's strategy of asking $p$. If player 1 uses strategy $x_1$, player 2's information set where he is asked to pay $p$ is not reached. However, at that information set, there is no action of player 2 that under all strongly consistent beliefs does at least as well as the excluded action of acceptance. The alternative of rejecting the offer and opening the blue box, say, and the alternative of rejecting it and tossing a coin are both worse than accepting the offer under the belief that, if player 1 asks $p$, then the money is in the yellow box. (This belief is strongly consistent with $x_1$, and reflects a hypothesis that player 1 deviated by asking $p$ only in case $Y$.) This conclusion contradicts the assumption that $X$ is an agnostic sequential polyequilibrium. The contradiction leaves $1/2$ as the only possible ASPE payoff. This payoff is obtained if player 1 asks $1/2$ (in both cases, $B$ and $Y$) and player 2 is willing to pay this price but would reject any higher price. It remains to show that this polystrategy is an ASPE.

The indicated polystrategy of player 2 does not specify a reaction to $p < 1/2$, and does so only partially for $p > 1/2$ by excluding acceptance. Put differently, the polystrategy excludes any strategy that instructs player 2 to (i) reject $p = 1/2$ (and then open the blue box, open the yellow box, or flip a coin), or (ii) accept at least one price $p > 1/2$. Given such an excluded strategy $x_2'$, let $x_2''$ be the strategy that differs from $x_2'$ only in that it instructs player 2 to (i) accept $p = 1/2$, and (ii) reject any price $p > 1/2$ acceptable by $x_2'$ and then flip a coin. For player 1's strategy $x_1$ of (always) asking $1/2$, if $U$ is player 2's information set where the price $p$ asked was indeed $1/2$, then the unique minimally sufficient set of actions is $\mathcal{A} = \emptyset$ and inequality (11) holds as equality for $i = 2$ as both sides are equal to $1/2$. If $U$ is an information set where $p > 1/2$ and $x_2'$ and $x_2''$ specify different reactions (respectively, acceptance and rejection followed by a coin flip), then a minimally sufficient set $\mathcal{A}$ consists of a single action, which is either asking price $p$ in case $B$ or doing so in case $Y$. With the former, $U$ is reached in case $B$ only, and in this case, strategy $x_2'$ yields player 2 the payoff $1 - p$ while $x_2''$ yields the higher (expected) payoff $1/2$. Therefore, (11) holds as strict inequality. The same is true for the second case, which proves that the polystrategy profile indicated at the end of the previous paragraph is indeed an ASPE.

## References

Blume, A., Heidhues, P. (2006). Private monitoring in auctions. *Journal of Economic Theory* 131, 179–211.

Cho, I.-K., Kreps, D.M. (1987). Signaling games and stable equilibria. *The Quarterly Journal of Economics* 102, 179–221.

Fudenberg, D., Tirole, J. (1991). Perfect Bayesian equilibrium and sequential equilibrium. *Journal of Economic Theory* 53, 236–260.

Hendon, E., Jacobsen, J., Sloth, B. (1996). The one-shot-deviation principle for sequential rationality. *Games and Economic Behavior* 12, 274–282.

Kohlberg, E., Mertens, J.-F. (1986). On the strategic stability of equilibria. *Econometrica* 54, 1003–1037.

Kreps, D.M., Ramey, G. (1987). Structural consistency, consistency, and sequential rationality. *Econometrica* 55, 1331–1348.

Kreps, D.M., Wilson, R. (1982). Sequential equilibria. *Econometrica* 50, 863–894.

Mas-Colell, A., Whinston, M.D., Green, J.R. (1995). *Microeconomic Theory*. Oxford University Press, Oxford, UK.

McLennan, A. (1985). Justifiable beliefs in sequential equilibrium. *Econometrica* 53, 889–904.

Milchtaich, I. (2019). Polyequilibrium. *Games and Economic Behavior* 113, 339–355.

Myerson, R.B. (1978). Refinements of the Nash equilibrium concept. *International Journal of Game Theory* 7, 73–80.

Myerson, R.B., Reny, P.J. (2015). Sequential equilibria of multi-stage games with infinite sets of types and actions. Mimeo.

Myerson, R.B., Reny, P.J. (2020). Perfect conditional ε-equilibria of multi-stage games with infinite sets of signals and actions. *Econometrica* 88, 495–531.

Perea, A. (2002). A note on the one-deviation property in extensive form games. *Games and Economic Behavior* 40, 322–338.

Selten, R. (1975). Reexamination of the perfectness concept for equilibrium points in extensive games. *International Journal of Game Theory* 4, 25–55.

Van Damme, E. (1984). A relation between perfect equilibria in extensive form games and proper equilibria in normal form games. *International Journal of Game Theory* 13, 1–13.

Van Damme, E. (1991). *Stability and Perfection of Nash Equilibria*, Second Edition. Springer-Verlag, Berlin.